# Profiling THz Beams With Off-Label Use of Infrared Microbolometric Cameras

Gabriel Nagamine,[1] Carlo Vicario,[2] Tariq Leinen,[1] Guy Matmon,[2] Marco Raffa,[3] Mattias Beck,[3] Giacomo Scalari,[3] Adrian L. Cavalieri,[1,2] Flavio Giorgianni[1,2]

*[1]Institute of Applied Physics, University of Bern, CH-3012 Bern, Switzerland*

*[2]Paul Scherrer Institute, CH-5232 Villigen-PSI, Switzerland*

*[3]Institute for Quantum Electronics, Department of Physics, ETH Zürich, CH-8093 Zurich, Switzerland*

**Abstract— Visualizing the spatial profile of light beams is essential for evaluating irradiance, characterizing beam quality, and achieving precise alignment. In the optical spectral range, this is readily performed using silicon-based CCD and CMOS cameras. In the terahertz (THz) range, however, it typically requires specialized detectors with prohibitive costs. Here, we show that an infrared (IR) camera can be used outside of its labeled specifications to achieve similar performance as a dedicated microbolometric THz camera, at under 1% of the THz camera's cost. We compared the cameras by characterizing THz beam profiles from two sources: a pulsed broadband THz beam produced through optical rectification in organic crystals, and a narrowband quasi-continuous-wave (quasi-CW) THz beam emitted by a quantum cascade laser. For the broadband THz radiation, the beam width measured by the two cameras differed by only ~ 6%, well within the pixel resolution limit, and in the narrowband quasi-CW case by just ~ 1.3%. Additionally, the IR camera exhibits a lower minimum detectable power (down to 1.5 THz) than the THz camera, while also maintaining a linear and polarization-independent responsivity. These results expand the applicability of conventional IR cameras to the THz range, suggesting that they will become routine tools for high-fidelity THz beam diagnostics and imaging in scientific and industrial applications.**

Beam profiling is a standard routine in optics, as it quantifies the spatial intensity distribution of light, enabling precise alignment in applications ranging from laser machining to imaging and spectroscopy [1]. From the infrared to ultraviolet range, beam profiling is routinely performed using semiconductor-based CCD/CMOS cameras [2]. In these cameras, photons in the electron-volt range excite electron–hole pairs to generate an appreciable electronic signal. However, such devices are not directly applicable in the terahertz (THz) regime, where the photon energies are comparable to thermal energy at room temperature. As a result, direct photon detection of THz radiation requires cryogenic cooling.

To overcome the need for cryogenic cooling, commercial manufacturers of THz cameras have explored different detection principles. For instance, companies like Terasense offer cameras where the THz radiation is detected by the direct excitation of plasma oscillations in a two-dimensional electron system. While this approach is highly effective in the sub-THz regime, these devices often rely on the use of resonant structures that limit the sensitivity at frequencies above 1 THz [3]. Alternatively, other technologies rely on thermal detection, where the incident radiation is absorbed and converted into heat, causing a measurable change in the detectors' physical properties [4]. This category includes pyroelectric arrays, offered by companies such as Ophir, which measure temperature-induced changes in electrical polarization [5]. However, as this method is only sensitive to temperature variations, it requires the use of schemes with mechanical optical choppers to image continuous-wave (CW) or quasi-CW sources. To address these limitations, microbolometric focal-plane arrays, offered by companies such as NEC or INO, are currently one of the most widely used sensors in state-of-the-art THz cameras. These detectors uniquely combine room-temperature operation with two-dimensional imaging, do not require optical chopping, and provide sufficient sensitivity for real-time imaging [6-8] of practical THz sources [3-5, 9].

Microbolometers, however, are not exclusive to the THz domain. The same thermal detection principle underpins inexpensive, mass-produced infrared (IR) thermal cameras, which benefit from CMOS-compatible fabrication, mature industrial processes, and large-scale production [10]. While THz and IR microbolometers exploit the same fundamental operating principles, they differ substantially in optical coupling strategies, absorber design, and pixel geometry, which are nominally optimized for distinct wavelength ranges and fabrication scales [3]. These engineering differences and

Corresponding Authors
E-mails: flavio.giorgianni@unibe.ch, gabriel.nagaminegomez@unibe.ch

production scale largely set the differences in cost and availability between THz and IR microbolometric cameras.

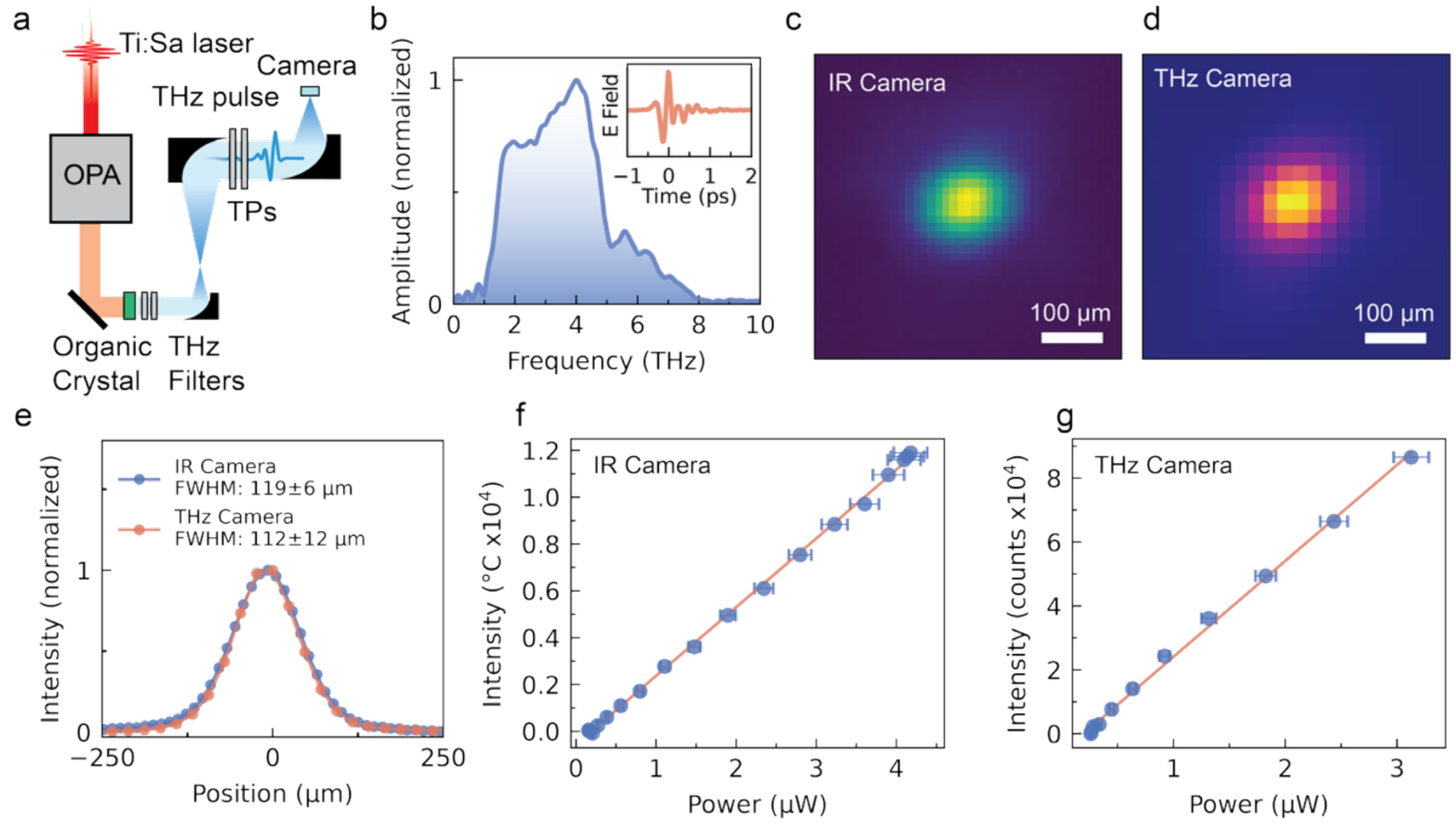

**Fig. 1 Comparison between THz and IR cameras for THz detection.** (a) Schematic of the THz beam profiling setup. The signal beam from an optical parametric amplifier (OPA) is optically rectified in a DSTMS organic crystal, generating broadband THz radiation (1-8 THz). To avoid leakage of the fundamental, three low-pass THz filters, two with cutoff frequencies of 20 THz, and the other with a cutoff frequency of 9 THz, are inserted in the THz beam. The THz beam is then focused on the cameras, after passing through two THz polarizers (TPs) to control the incident power. (b) THz amplitude spectrum obtained through electro-optic sampling of the THz electric field in the time domain (shown in the inset). (c-d) Image of the focused THz beam profile on the IR and THz cameras. Note that distinct colormaps are used to visually differentiate the two devices. (e) Comparison of horizontal cross-cuts obtained from the measured intensity profiles in (c, d). (f-g) Linearity of the signal strength as a function of THz irradiance. The *x* axes correspond to the average power, while the *y* axes report the raw integrated signal given by the cameras. Note that for the IR camera, the reported temperature corresponds to the spatially integrated pixel signal (sum over the beam area) and does not represent an actual thermodynamic temperature, whereas the THz camera output is given in counts.

If IR microbolometric cameras could be proven useful in the THz – beyond their design specifications – they would be an attractive alternative to dedicated THz cameras. In this direction, a previous proof-of-concept study [11] has suggested that the dominant differences in detection sensitivity due to wavelength (THz vs. IR) are not intrinsic to the microbolometric sensing mechanism. Rather, they appear to arise primarily from different analog-to-digital conversion electronics and from the specific implementation of dynamic background-subtraction algorithms. However, the study described in Ref. [11] was limited to the spectral content below 2.5 THz, and key performance parameters, such as linearity, responsivity, and spectral sensitivity, were not fully explored.

In this paper, the linearity, spectral and polarization-dependent sensitivities, noise spectral density, and beam profiling capabilities of two commercially available microbolometric cameras that employ vanadium oxide (VOx) as the thermoresistive sensing material are evaluated and compared. Similar performance between the HIKMICRO Mini2PlusV2 IR camera (sold at ~$249) and the NEC IR/V-T0831 THz camera (sold at ~$30,000) is observed under both broadband THz and narrowband THz illumination. These results indicate that IR microbolometric cameras can indeed be used to accurately capture the THz radiation, offering an affordable solution for THz imaging and beam profiling for research and industrial applications.

The first set of comparative measurements is performed using a focused broadband THz beam generated by laser-driven optical rectification in organic crystals. It should be noted that for clarity, the HIKMICRO Mini2PlusV2 camera will be referred to as the “IR camera” and the NEC IR/V-T0831 as the “THz camera”. Additionally, in all measurements with the IR camera, the IR collection lens was removed to allow direct illumination of the microbolometric array. This is required as common IR lens materials, such as doped germanium and chalcogenide glasses, do not transmit THz radiation [11-13].

A sketch of the experimental setup used to characterize the cameras under broadband THz illumination is shown in Fig. 1a. A 1 kHz Ti:Sapphire femtosecond laser system coupled to an optical parametric amplifier (OPA) delivers a 1.35 µm pulse, which is optically rectified in a DSTMS [14] or PNPA [15] organic crystal to generate collimated broadband THz radiation. Fig. 1b shows the THz amplitude spectrum, which is centered around 4 THz and extends between 0.3 and 8 THz. The spectrum is obtained from the Fourier amplitude of the THz pulse waveform (inset of Fig. 1b), which is measured by electro-optic sampling (EOS), as described in Section S1 of the Supplementary Material. After generation, the THz beam is isolated using a set of two 20 THz and one 9 THz low-pass multi-mesh filters, which have an extinction ratio exceeding $10^6$ from the mid-infrared to the near-infrared range (see Supplementary Figure S6). We note that using multiple thermally isolated filters is important for suppressing residual pump leakage and the thermal infrared background generated by laser-induced heating, thereby ensuring that the signal has only THz contributions (see discussion in section S4.3 of the Supplementary Materials). The beam then propagates through a set of polarizers to adjust the transmitted power. Finally, the beam is expanded and tightly focused (numerical aperture, NA~0.36) onto the cameras using three off-axis parabolic mirrors. Additional experimental details, including IR camera data handling, background subtraction procedure, and setup specifications, are reported in Section S1 of the Supplementary Material. In Section S4 of the Supplementary Material, we present a series of control experiments that rule out contributions from pump leakage and laser-induced thermal radiation, ensuring that the measured signals throughout this work originate from THz radiation.

Images of the focused THz beam at an average intensity of 7 mW/cm², acquired using the IR and THz cameras, are shown in Fig. 1c and Fig. 1d, respectively. Both images are shown after background subtraction and normalization. Horizontal cross-cuts of the intensity profiles are shown in Fig. 1e. Qualitatively, the two cameras capture nearly identical spatial beam distributions.

Quantitatively, the full-width at half-maximum (FWHM) is $119 \pm 6$ μm for the IR camera and $112 \pm 12$ μm for the THz camera, corresponding to a relative difference of only ~6%. The uncertainty in the FWHM is estimated as half of the pixel pitch of each detector (12 μm for the IR camera and 23.5 μm for the THz camera), indicating that the observed discrepancy is well within the spatial sampling limit of the sensors. The close agreement with the benchmark THz camera confirms that the IR camera can accurately resolve the THz beam profile.

In general, THz cameras employ larger pixels to increase the absorber area and enhance responsivity at long wavelengths [7, 16]. While both pixel pitches are smaller than the THz wavelength (~74 μm at 4 THz), finer sampling provided by the smaller pixels of the IR camera can, in principle, enable improved spatial resolution.

To verify the polarization dependence of the detector response, measurements were performed using linearly polarized THz radiation with horizontal and vertical polarization orientations (see Section S1 of the Supplementary Material). For both cameras, the relative deviation in the integrated signal between polarization states was approximately 0.3%, confirming the isotropic response of both detectors (Supplementary Table 1). While the thermal detection mechanism of microbolometers is, in principle, insensitive to the polarization of the incident radiation, the pixel geometry—particularly in designs incorporating antenna or resonant coupling structures—can introduce anisotropic or polarization-dependent responses. The negligible differences observed here indicate that the internal detector architecture does not introduce measurable polarization-dependent sensitivity in the investigated cameras. This polarization-insensitive response is particularly desirable for beam profiling and imaging applications, where the polarization state of the incident THz radiation may vary or be unknown.

The linearity in the response of the IR camera is also compared to the benchmark THz camera. Figs. 1f and 1g show the integrated image intensity as a function of the total incident THz average power (see Supplementary Material for details). The THz power is varied using two THz polarizers in series before the imagers. Both cameras exhibit nearly perfect linear response over the full range of intensities. A linear regression yields a slope of $2.9 \times 10^3$ °C/μW for the IR camera and $3.0 \times 10^4$ counts/μW for the THz camera, with relative standard error of the slope of 0.7 and 1.7%, respectively. Because the IR camera is operating outside its labeled specifications, its software natively reports the detected radiation as a temperature reading (°C), whereas the dedicated THz camera outputs raw intensity counts. In the present off-label use of the IR camera for THz detection, the reported temperature therefore serves as a proportional measure of incident THz intensity.

A key parameter in validating the use of infrared (IR) cameras for beam profiling is the spectral range over which these devices operate reliably. To determine the operational range, the responsivity $R$ and the sensitivity of both detectors were evaluated as a function of the incident THz frequency. To vary the incident THz spectral content, a set of low-pass THz filters with progressively lower cut-off frequencies was used.

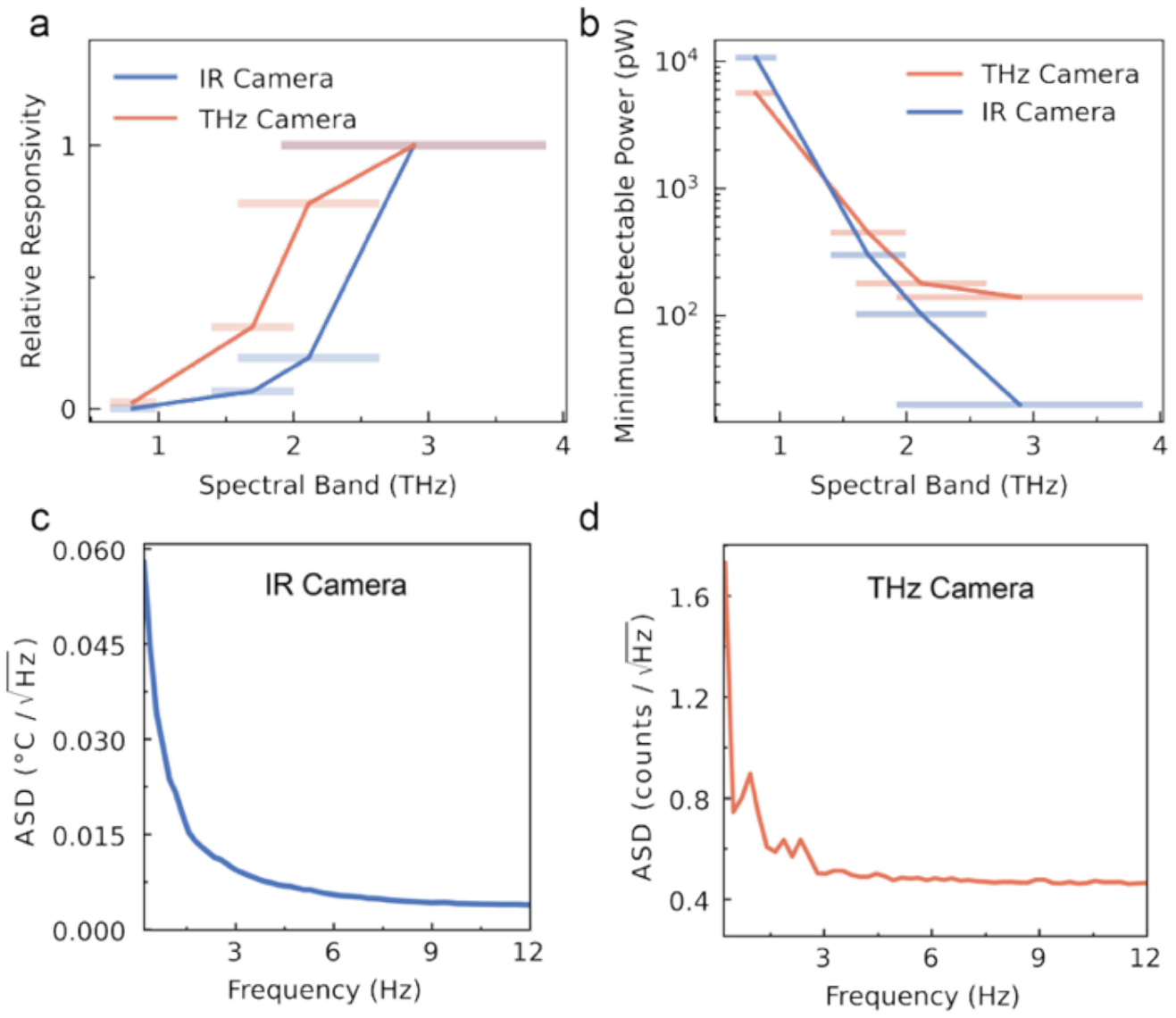

**Fig. 2 Spectral dependence of cameras' responsivity, sensitivity, and noise analysis.** Relative responsivity (a) and minimum detectable power (MDP) (b) as a function of the incident frequency band. The relative responsivity is defined as the ratio between the camera's responsivity with the low-pass filters and that with the full incident spectrum. The horizontal bars in (a) and (b) denote the spectrum-weighted standard deviation, and they are centered at the spectrum-weighted mean. Amplitude spectral density (ASD) of the noise for the IR (c) and THz (d) cameras. The frequency axis in (c) and (d)

corresponds to the temporal noise spectrum derived from time-series measurements.

The responsivity for a given spectral band was calculated as the spatially integrated camera signal divided by the corresponding incident average power (see Section S1 of the Supplementary Material). Because the two detectors produce integrated output signals in different units (temperature in °C for the IR camera and counts for the THz camera), the *relative responsivity*, defined as the responsivity for a given spectral band normalized to the value measured under full broadband excitation, is the most relevant quantity. Fig. 2a shows the relative responsivity of the IR and THz cameras as a function of the incident spectral band. The center frequency $\mu_w$ and weighted standard deviation $\sigma_w$ of each spectral band were extracted from electro-optic sampling measurements of the incident THz spectrum (see Supplementary Material) and are depicted in Fig. 2a by the center and width of the horizontal bars.

For the THz camera, the relative responsivity falls to 50% at 1.5 THz, whereas for the IR camera, the corresponding 50% drop occurs at a higher frequency, 2.5 THz, with a steeper decline. A reduction in responsivity in the lower-THz regime is a well-known limitation of microbolometer-based detectors [16, 17], and improving performance in this spectral region remains an active area of research [7, 16]. A widely adopted approach to improve responsivity at long THz wavelengths is to increase the effective pixel area so that the absorber captures a larger fraction of the subwavelength electromagnetic field [7, 16]. Consistent with this design strategy, the difference in relative responsivity between the cameras at lower THz frequencies correlates with the larger pixel pitch of the benchmark THz camera, which is approximately twice that of the IR camera (23.5 μm versus 12 μm).

The THz frequency-dependent sensitivity was calculated as the minimum detectable power (MDP), defined as:

$$MDP = \frac{\sigma}{R} \qquad \text{Eq. 1}$$

where $\sigma$ is the temporal standard deviation of the detector noise, obtained from a 300-frame time-series measurement without incident radiation, and $R$ is the responsivity. Fig. 2b shows the MDP of both detectors across the incident spectral bands. Due to the frequency-dependent responsivity, the MDP increases in the region around ∼1 THz by approximately three orders of magnitude for the IR camera and two orders of magnitude for the THz camera.

Under full-spectral excitation (center frequency $\mu_w = 2.9\ THz$), the measured MDPs were 20 pW for the IR camera and 140 pW for the THz camera. We note that a direct comparison between these values must be treated with caution, as the cameras have different frame rates (30 Hz for the THz camera and 25 Hz for the IR camera), different readout electronics, and different signal-processing algorithms. Nonetheless, this characterization shows that the IR camera exhibited lower MDPs for the higher-frequency components of the organic crystal emission spectrum, whereas the THz camera performed better at frequencies below 1.5 THz.

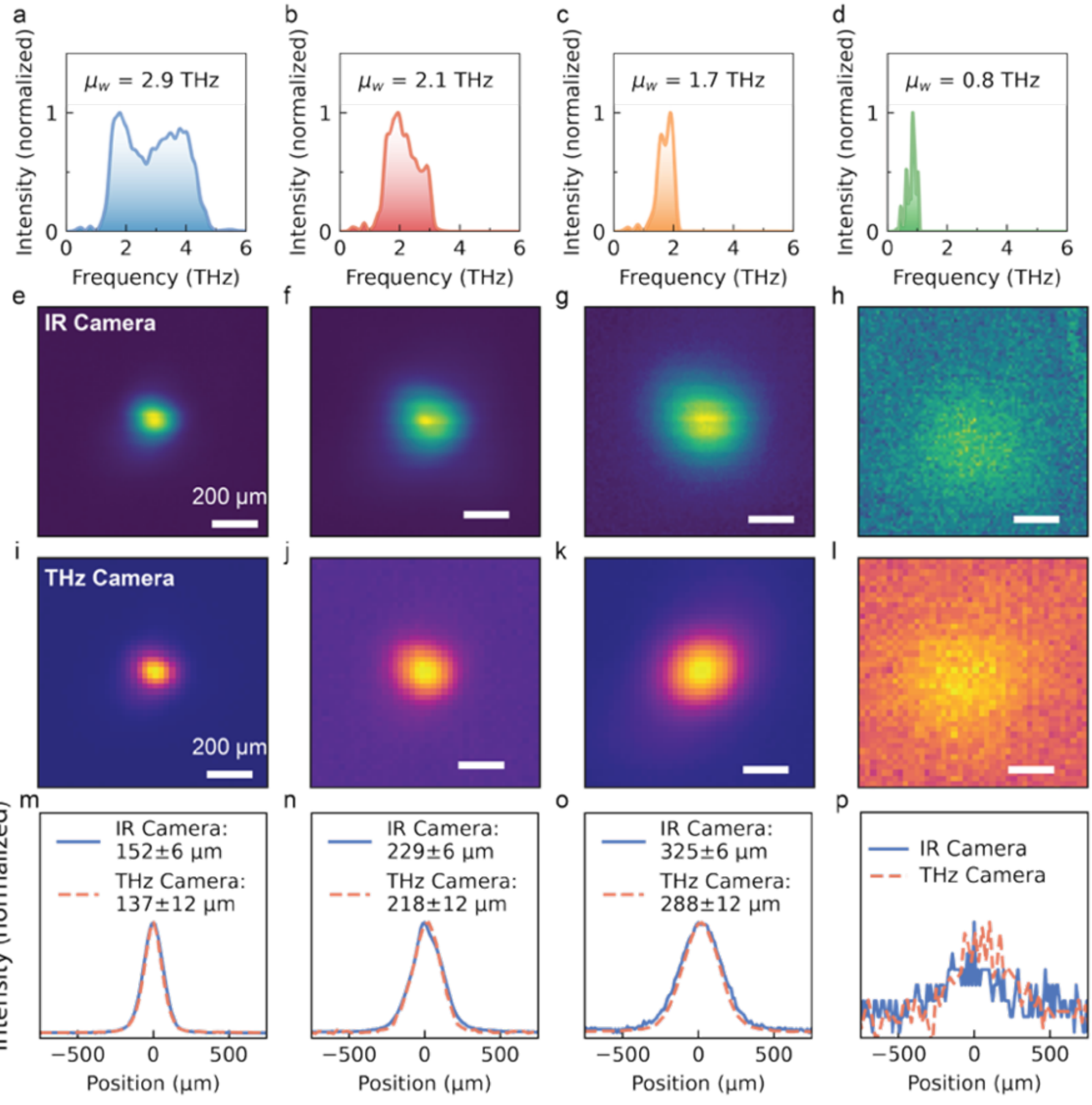

**Fig. 3 Terahertz beam focus profiles across different THz spectral bands using THz lowpass filters.** (a-d) Amplitude spectrum of the transmitted THz beam. The frequency bands were controlled using low-pass THz filters with cutoff frequencies of 9.0, 3.0, 2.0, and 1.0 THz, respectively (see Supplementary Material S1). The central frequency (weighted mean) calculated from the measured THz waveform is indicated in each panel. Images (e-l) and cross-cuts (m-p) of the THz beam profile recorded with the IR (e-h) and THz (i-l) cameras. All scale bars correspond to 200 μm.

While the MDP provides a practical detection limit, it relies on a temporal standard deviation that integrates the noise across the detector's entire operational bandwidth. As a result, the MDP is inherently dependent on the effective measurement bandwidth, which is determined by camera parameters such as integration time or filtering characteristics of readout electronics. To provide a more comprehensive characterization of the noise floor (and to enable a more direct comparison between different detection systems), the amplitude spectral density (ASD) of the output noise (see section S1 of the Supplementary Material) is calculated. Figs. 2c and d show that both detectors have different noise profiles. While the noise from the IR camera is dominated by 1/f (pink) noise, the noise from the THz camera contains peaks in the region below 3 Hz. In contrast, for frequencies above 3 Hz, the THz camera ASD is flatter, indicating a more stable noise floor.

As microbolometric detectors exhibit a frequency-dependent cutoff in their spectral responsivity (and sensitivity), beam-profiling measurements performed with broadband THz radiation may, in principle, underestimate the true focal spot size by preferentially weighting higher-frequency components.

To isolate the magnitude of this effect (for profiling broadband THz emission from organic crystals), we performed numerical simulations using the experimentally measured spectral responsivities of both cameras. The purpose of this calculation is not to reproduce the absolute measured focal spot size—which depends on optical alignment, aberrations, and the wavelength-dependent initial beam waist—but rather to quantify the bias introduced solely by detector spectral weighting. Starting from the relative responsivity curves shown in Fig. 2a, continuous responsivity functions were constructed and applied to the measured DSTMS emission spectrum. Each spectral component was then propagated through the focusing geometry using a Gaussian beam model, with the same optical input assumed for both detectors. In this way, any difference in the reconstructed beam profile arises exclusively from the detectors' spectral responsivity. Full details of the numerical procedure are provided in Section S3 of the Supplementary Material.

Using a spectrally flat (ideal) responsivity as a reference, the reconstructed focal width is 120 μm. When the measured spectral responsivities of the THz and IR cameras are applied, the corresponding reconstructed widths become 116 μm and 112 μm, respectively, corresponding to deviations below 7% relative to the flat-responsivity case. These small relative differences indicate that spectral roll-off alone does not introduce a significant distortion in the reconstructed beam width for broadband spectra dominated by frequencies above 2 THz.

To further investigate the dependence of the observed beam profile on frequency, measurements are made as the incident THz spectrum is systematically shifted toward lower frequencies using a series of low-pass THz filters. The resulting beam profiles for the different spectral bands are shown in Fig. 3. The amplitude spectra for the different filter settings are displayed in the top row, and the corresponding measurements made with the IR and THz cameras for that particular spectral range are shown below. For the lowest-frequency band centered at 0.8 THz (Figs. 3d, 3h, and 3l), the DSTMS emitter was replaced with a PNPA organic crystal [15], which has a higher THz conversion efficiency below 1.5 THz.

As the spectral content is shifted toward lower frequencies, the effective beam size increases due to the larger diffraction-limited spot size at longer wavelengths. Moreover, at higher frequencies, the acquired profiles exhibit enhanced signal-to-noise ratios, yielding well-resolved intensity distributions. The performance degrades for both cameras at the lowest frequency ($\mu_w$ = 0.8 THz, Figs. 3h and 3l), where the power is lower (see Supplementary Table 2). Combined with the reduced responsivity of both cameras at these frequencies, this leads to a decrease in the signal-to-noise ratio.

A quantitative comparison of the measured beam widths clarifies the impact of spectral effects on beam profiling. For spectral bands centered at $\mu_w = 2.9$ THz and $\mu_w = 2.1$ THz, the FWHM measured by the IR and THz cameras agrees within the spatial sampling uncertainty set by the respective pixel pitches. In contrast, for the spectral band centered at $\mu_w = 1.7$ THz, the IR camera yields a beam diameter that is approximately 11% larger than that measured with the THz camera, exceeding the expected sampling-related uncertainty.

Notably, this discrepancy cannot be explained by the reduced responsivity at lower frequencies, as this would preferentially weight higher-frequency components and lead to a smaller, rather than larger, measured beam diameter. The observed broadening when $\mu_w = 1.7$ THz may instead be attributed to a modest contribution from pixel-to-pixel cross-talk in the IR camera [18]. Such thermal or radiative cross-talk, arising from non-perfect thermal or electromagnetic isolation between neighboring pixels, may become more pronounced when operating near the lower-frequency limit of the device, where the responsivity is lower. Importantly, a discrepancy exceeding the spatial sampling uncertainty (set by the respective pixel pitches) was observed only for spectral bands below 2 THz, and therefore does not affect the overall ability of the IR camera to accurately reconstruct broadband THz beams. At the lowest spectral band ($\mu_w = 0.8$ THz), the detected signal is substantially reduced due to the diminished responsivity of both cameras, resulting in a low signal-to-noise ratio. At this frequency, a reliable extraction of the FWHM is not possible, and no quantitative comparison of beam diameters is reported.

While the previous measurements are made to characterize the cameras under broadband pulsed THz excitation, many practical applications, especially imaging, rely on CW and quasi-CW THz sources. Therefore, in a second set of measurements, the performance of the IR camera under quasi-CW conditions was evaluated. An on-chip quantum cascade laser (QCL) providing a stable, multimode output at approximately 3 THz (bandwidth: ~250 GHz FWHM) was used for these tests. More details on the QCL, including emission spectrum and operation characteristics, can be found in ref. [19] where the same QCL platform used in this work is described and characterized.

The QCL itself is pulsed at a 200 kHz repetition rate, with a 10% duty cycle, and this pulse train is then gated at 105 Hz using a 50% duty cycle to match the camera's integration time. A schematic of the experiment and the measured beam profiles are shown in Fig. 4. The profiles were acquired at an incident THz average power attenuated to below 2 µW. Consistent with the results obtained with the broadband THz excitation, the IR camera produced a beam profile of the QCL that closely matched the profile from the THz camera. Specifically, the horizontal FWHM spot size measured with the IR camera was $141 \pm 6\ \mu m$, compared to $139 \pm 12\ \mu m$ for the THz camera. Near-perfect agreement under these conditions further confirms the suitability of the IR microbolometer for accurate THz beam profiling under both pulsed broadband and quasi-CW narrowband excitation.

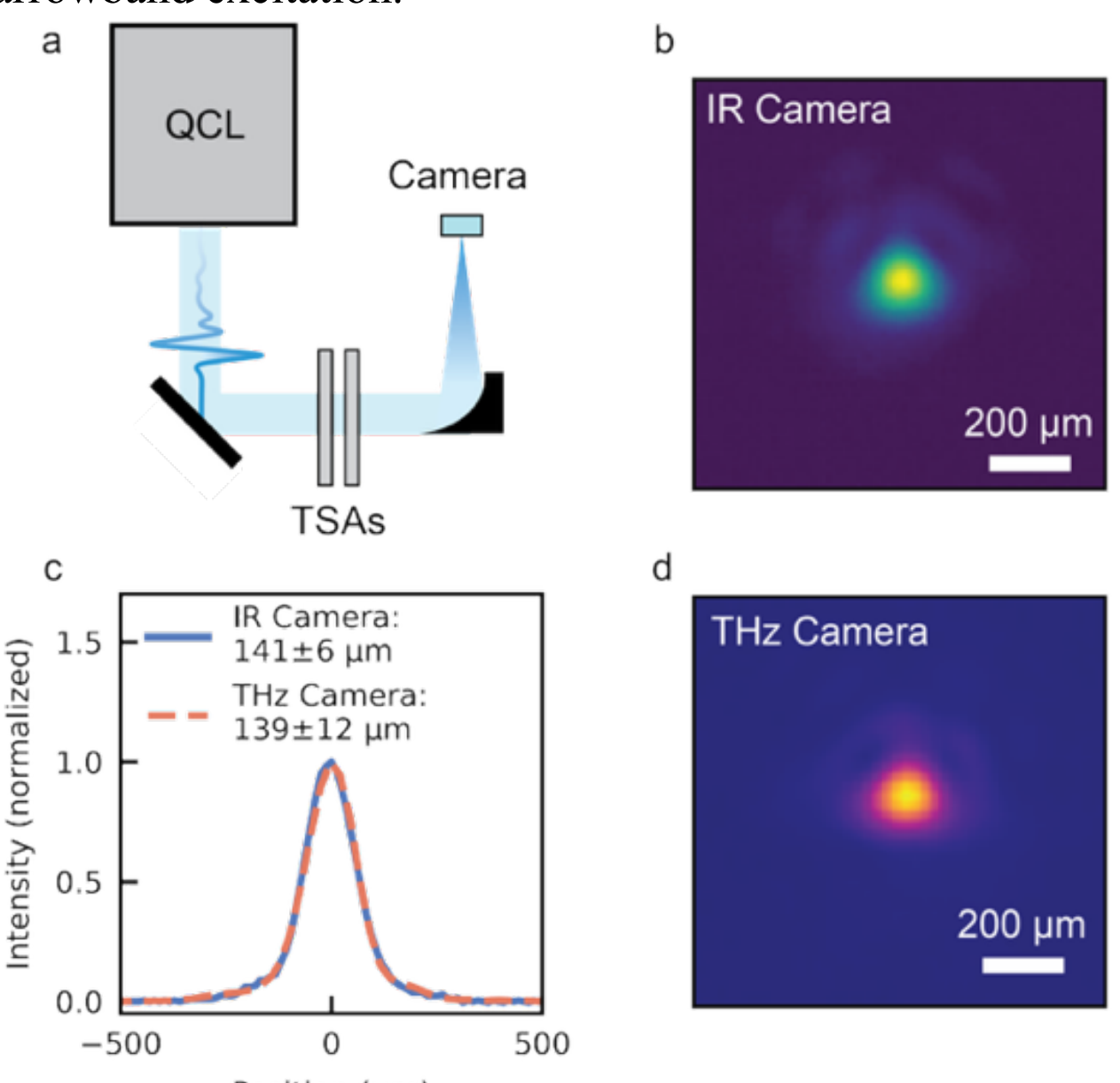

**Fig. 4 Quasi-Continuous-Wave THz imaging with a quantum cascade laser (QCL).** (a) Illustration of the experimental setup used to test the cameras under a quasi-CW regime. A QCL was used to generate a beam centered at 3 THz, which was focused onto the camera's detector. The power was controlled with THz silicon attenuators (TSAs) placed in series. (b) Image of intensity profile in the IR camera. (c) Horizontal cross-cut of the beam profiles from (b) and (d), after background subtraction. (d) Beam image recorded with the THz camera. Scale bars in (b) and (d) are 200 µm.

In summary, relatively inexpensive, IR microbolometric cameras can be used for beam profiling with similar performance to that of a specialized THz camera. For broadband THz radiation from organic crystals, the IR camera reproduced the beam profile with deviations below ~6%. Notably, this discrepancy corresponds to less than the sensor's 12 µm pixel pitch, suggesting that the observed difference is limited primarily by spatial sampling rather than detector sensitivity. Furthermore, simulations incorporating the measured spectral responsivity indicate that the low-frequency roll-off introduces only minor distortions (<7%). Measurements using a 3-THz on-chip QCL provide additional agreement between the two systems, yielding spot-size differences of only ~1.3%. Overall, this work shows that off-the-shelf IR microbolometric cameras can be used as reliable and accurate instruments for THz beam diagnostics, establishing these cameras as a standard tool for routine alignment and imaging in terahertz science and technology.

## Acknowledgment

The authors express their gratitude to the PSI-LNO GL group for their support during the experiments and to Willy Stehl for the discussions.

**Supplementary Material for**

# Profiling THz Beams With Off-Label Use of Infrared Microbolometric Cameras

Gabriel Nagamine,[1] Carlo Vicario,[2] Tariq Leinen,[1] Guy Matmon,[2] Marco Raffa,[3] Mattias Beck,[3] Giacomo Scalari,[3] Adrian L. Cavalieri,[1,2] Flavio Giorgianni[1,2]

*[1]Institute of Applied Physics, University of Bern, CH-3012 Bern, Switzerland*
*[2]Paul Scherrer Institute, CH-5232 Villigen-PSI, Switzerland*
*[3]ETH Zürich, CH-8093 Zurich, Switzerland*

## S1. Experimental Methods

**Experimental Setup for Beam Profiling in Pulsed Excitation.** The experimental setup for the beam-profiling measurements is illustrated in Figure 1a of the main text. A 1 kHz Ti:Sapphire laser, coupled with an optical parametric amplifier, was used to pump an organic crystal, generating collimated THz radiation. The pump wavelength was ~ 1350 nm, for an efficient THz generation through optical rectification. In this study, all measurements in the pulsed regime were performed by pumping a DSTMS organic crystal,[S1] except for the measurements shown in Figures 3d, h, and l of the main text, where we used PNPA[S2] due to its enhanced emission at lower THz frequencies.

To avoid leakage from the pump or parasitic black-body radiation generated by laser-induced heating, a set of two THz low-pass filters (QMC Instruments, multi-mesh filters) with a cutoff frequency of 20 THz, combined with one filter with a cutoff frequency of 9 THz, was positioned right after the organic crystal (see Fig S6 for the transmission spectrum). Fourier-transform infrared spectroscopy measurements indicate that these filters exhibit an extinction ratio exceeding $10^6$ across the mid-infrared to near-infrared range. A particularly important characteristic of this setup is the use of more than one thermally decoupled optical element. So, even though the organic crystal or the first filter can warm up due to the incident laser, the subsequent filters can remove the generated blackbody radiation (see discussion in section

---

Corresponding Authors
E-mails: flavio.giorgianni@unibe.ch, gabriel.nagaminegomez@unibe.ch

S4.3).

To generate a tight focus, the beam was expanded and then focused using a series of unprotected gold off-axis parabolic mirrors. The focal distances of the parabolic mirrors were, in sequence: 1, 6, and 2 inches.

The focused beam was then imaged by a Mini2PlusV2 infra-red (IR) camera from HIKMICRO and by a T0831 THz camera from NEC. For simplicity, in the main text, we refer to these cameras as “IR” and “THz” cameras, respectively.

**THz Power Measurements.** To measure power, a THZ5I-BL-BNC from Gentec was used. For the responsivity measurements, the power was controlled by varying the angle between two THz polarizers in series (TYDEX, polypropylene film-based polarizers).

**IR Camera Data Extraction and Background Subtraction.** The IR camera data were stored as radiometric files, containing the full per-pixel temperature information in addition to the RGB visualization. The temperature matrix was extracted using the analysis software provided by the manufacturer at the native spatial resolution of the detector array. Background correction was performed by acquiring two consecutive images, alternatively blocking and unblocking the pump beam, incident on the organic crystal. The background was then subtracted from the signal in a post-processing step using custom-written Python scripts.

**Determination of the Spatial THz Beam FWHM.** The spatial full width at half maximum (FWHM) of the focused THz beam was extracted from one-dimensional horizontal intensity profiles obtained from the beam images. For each profile, the half-maximum level was defined as half of the peak intensity of the spatial distribution. The positions at which the intensity crossed this level were determined using linear interpolation between adjacent data points surrounding each crossing. The FWHM was then calculated as the distance between the left and right interpolated half-maximum positions.

**Fourier-transform infrared (FTIR) measurements.** The spectral transmittance of the multi-mesh THz low-pass filters and the THz polarizer was characterized from the THz to the near-infrared range by Fourier-transform infrared spectroscopy using a Bruker IFS125HR spectrometer at the QPS laboratory of the Paul Scherrer Institute (see Figures S6 and S8). Measurements in the mid-infrared and near-infrared spectral ranges were carried out in air, while those in the THz/far-infrared range were performed under vacuum.

**Characterization of the Polarization Dependence of the Cameras’ Response.** To assess the isotropic response of the detectors, we have performed measurements with both cameras in horizontal and vertical orientations, while the THz light remained vertically polarized in relation to the laboratory frame. The incident THz power and linear polarization were kept constant for both configurations, and the full THz beam profile was recorded in each case. The integrated signal was then calculated for each orientation. The results are summarized in the Supplementary Table 1.

**Supplementary Table 1.** Summary of results for isotropic response characterization.

| | Vertical Orientation Integrated Signal | Horizontal Orientation Integrated Signal | Relative Signal Difference (%) |
|---|---|---|---|
| IR Camera | 5171 °C | 5156 °C | 0.3 |
| THz Camera | 75049 counts | 74832 counts | 0.3 |

**Computation of the Amplitude Spectral Density.** The amplitude spectral density (ASD) of the noise floor was obtained from the square root of the power spectral density (PSD) function. The PSD was calculated from a time series containing 300 frames, with the camera shutter blocked. The calculation was performed using the Welch's method[S5] implemented via the scipy signal[S6] library. The resulting PSD function was obtained after averaging over a 100x100-pixel block. Figure 2c and d from the main text show the obtained ASD functions for the IR and THz cameras, respectively.

**Quasi-continuous-Wave Excitation with a Quantum Cascade Laser.** For a quasi-continuous-wave illumination (CW), a specifically designed on-chip quantum cascade laser (QCL) was used to generate THz radiation at a frequency of 3 THz with 250 GHz bandwidth FWHM[S7]. The on-chip QCL was cooled at temperatures below 20 K, and it was driven at a 200 kHz repetition rate, with a 10% duty cycle. The emitted radiation was collected with an off-axis parabolic mirror and then refocused on the detector. A 6 THz low-pass filter was used to cut the ambient background. This micro-pulse train was then gated at 105 Hz using a 50% duty cycle. The incident power on the detector was < 2 μW, corresponding to a peak irradiance <13 mW/$cm^2$.

**Spectral Band Control Using Low-Pass THz Filter.** To vary the incident spectral band, a series of low-pass THz filters supplied by QMC Instruments Ltd. was used. For the spectral characterization, we performed electrooptical sampling (EOS).[S8] The experimental setup for EOS characterization is shown in Figure S1a. The obtained THz field in time is shown in Figure S1 b-e, while the respective fields in the frequency domain are shown in Figure S2. The cutoff frequencies for the THz filters were 9, 3, 2, and 1 THz. We identify each field based on the weighted mean ($\mu_w$) as done in the main text. For all fields, a DSTMS[S1]crystal was used to generate THz, except for the one centered at 0.8 THz, where we used PNPA [S2] for more efficient THz generation.

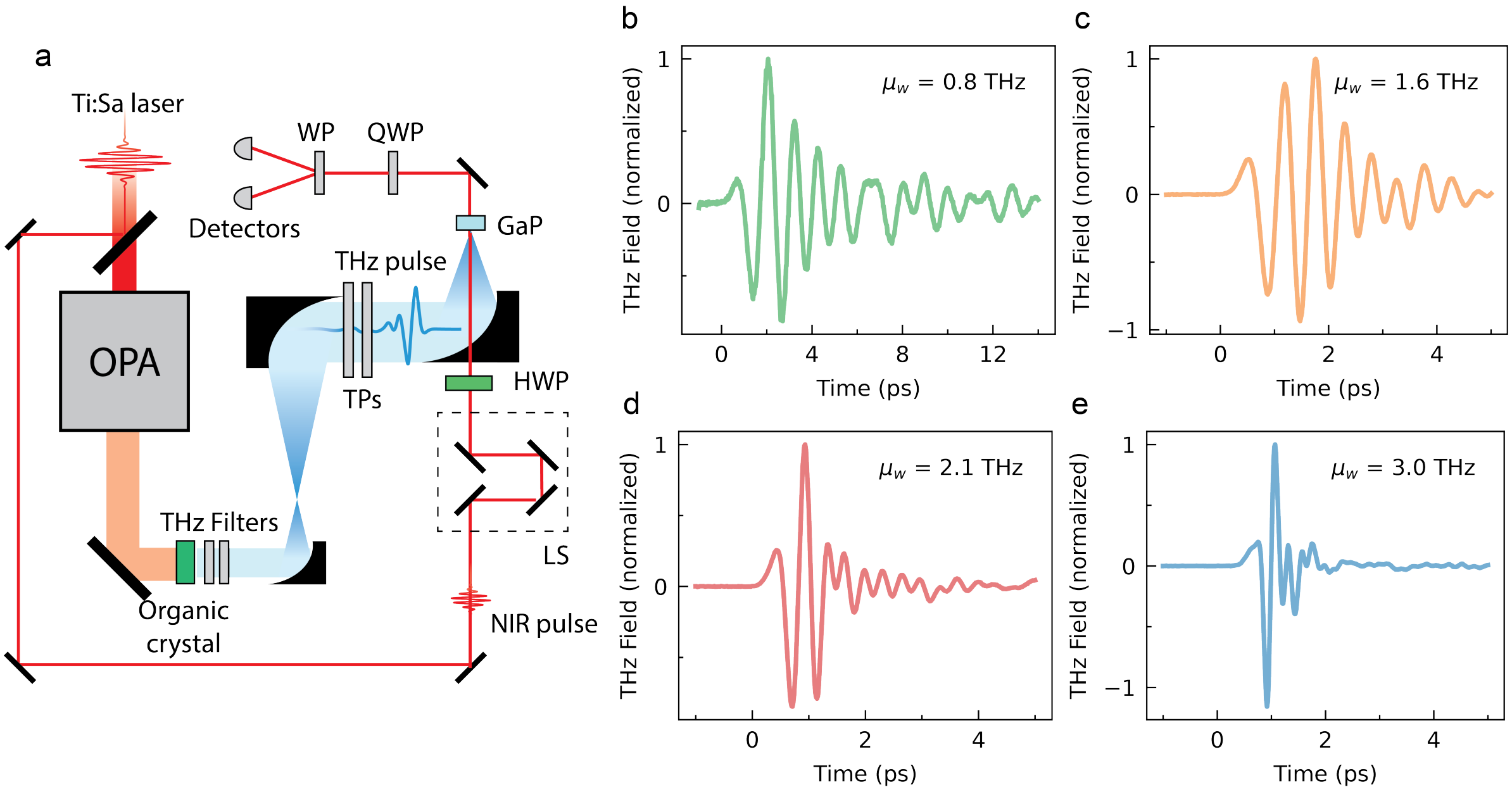

***Figure S1. Electro Optical Sampling for THz Field Characterization.*** *(a) Illustration of the setup used for electro-optical sampling (EOS). LS: linear stage, HWP: half-wave plate, TPs: THz polarizers, QWP: quarter-wave plate, WP: Wollaston prism. (b-e) THz field in the time domain, obtained from EOS. The timetraces were obtained using low-pass THz filters with cutoff frequencies of 1 THz (b), 2 THz (c), 3 THz (d), and 9 THz (e). The panels show the weighted mean frequency ($\mu_w$), obtained from the data in the frequency domain.*

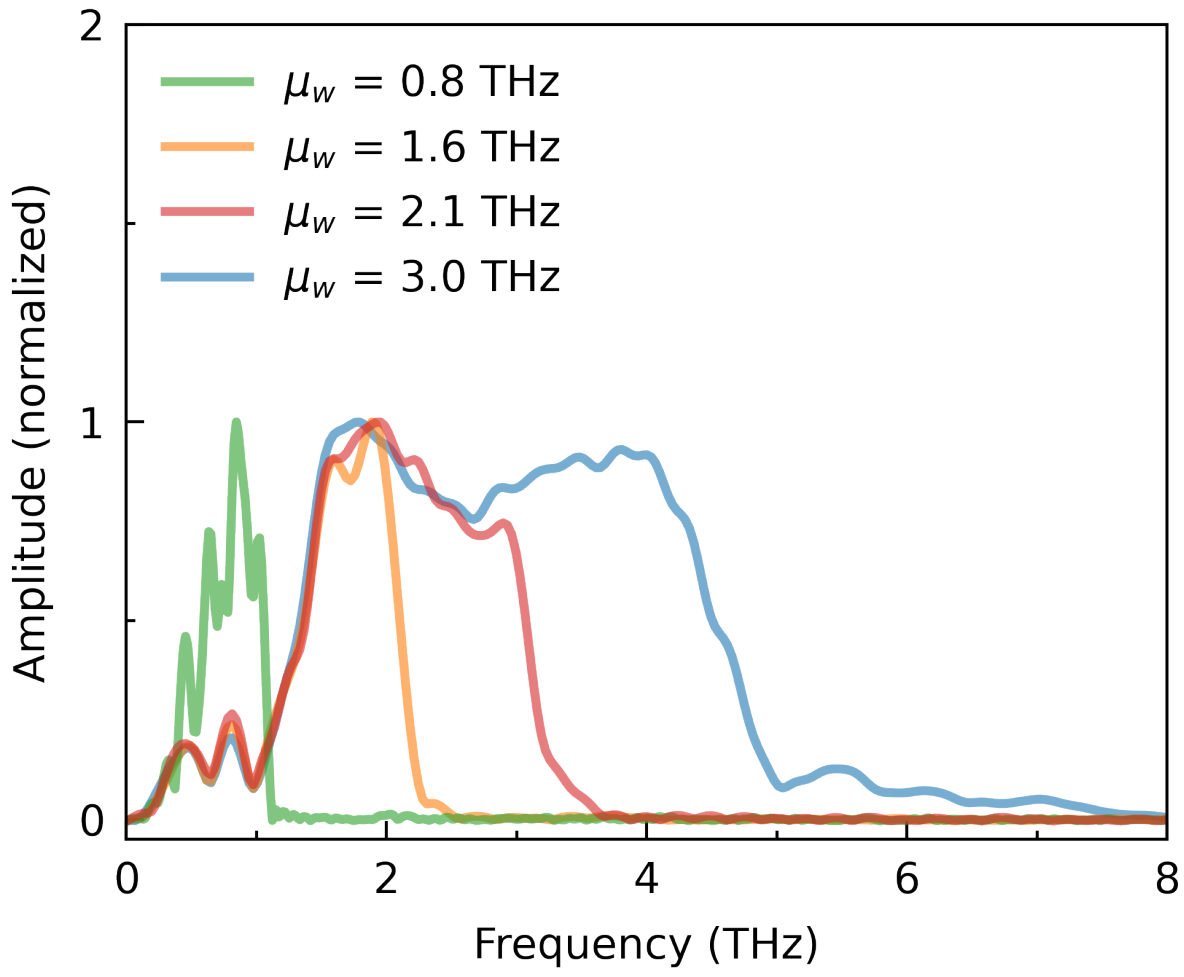

***Figure S2. THz Spectra Varying Incident Frequency Band****. The spectra were obtained from a fast Fourier transform of the data shown in Figure S1(b-e). To control the incident frequency band, we used low-pass THz filters with cutoff frequencies of 1 THz, 2 THz, 3 THz, and 9 THz, generating spectra with weighted means of 0.8, 1.6, 2.1, and 3.0 THz, respectively.*

## S2. Calculation of the Relative Responsivity

The responsivity at a given center frequency $f$ can be calculated by the following equation:

$$R_f = \frac{S_f}{P_f}, \quad \text{S1}$$

where $S$ and $P$ refer to the measured signal and power, respectively. The signal was measured as the integrated intensity over the beam profile, and the parameters used for the calculation are listed in Supplementary Table 2. The normalized responsivity was calculated by dividing $R_f$ by the responsivity obtained with full spectral excitation (weighted mean frequency of 2.9 THz).

**Supplementary Table 2.** Parameters used for calculating the responsivity.

| Weighted mean frequency (THz) | IR camera integrated signal (°C) | THz camera integrated signal (**counts**) | Incident Power (μW) |
| --- | --- | --- | --- |
| 0.8 | 40.4 | 8606 | 9.05 |
| 1.7 | 3325.5 | 249028 | 20.9 |
| 2.1 | 460.8 | 29638 | 1.0 |
| 2.9 | 8990.9 | 143760 | 3.8 |

## S3. Simulation of Beam-Profile Distortion from Spectrally-Dependent Responsivity

To quantify the distortions caused by the spectrally-dependent responsivity on beam profiling, we first extract a continuous responsivity function from the data shown in Figure 2 of the main text. To do so, we interpolated the data using a cubic Hermite interpolator (PchipInterpolator from SciPy[S6]). Then, the square root of the interpolated responsivity was applied to the DSTMS THz emission spectrum (blue curve from Figure S2) to generate a responsivity-weighted electric field spectrum. The extracted square root of the interpolated responsivity for both cameras is shown in Figure S3.

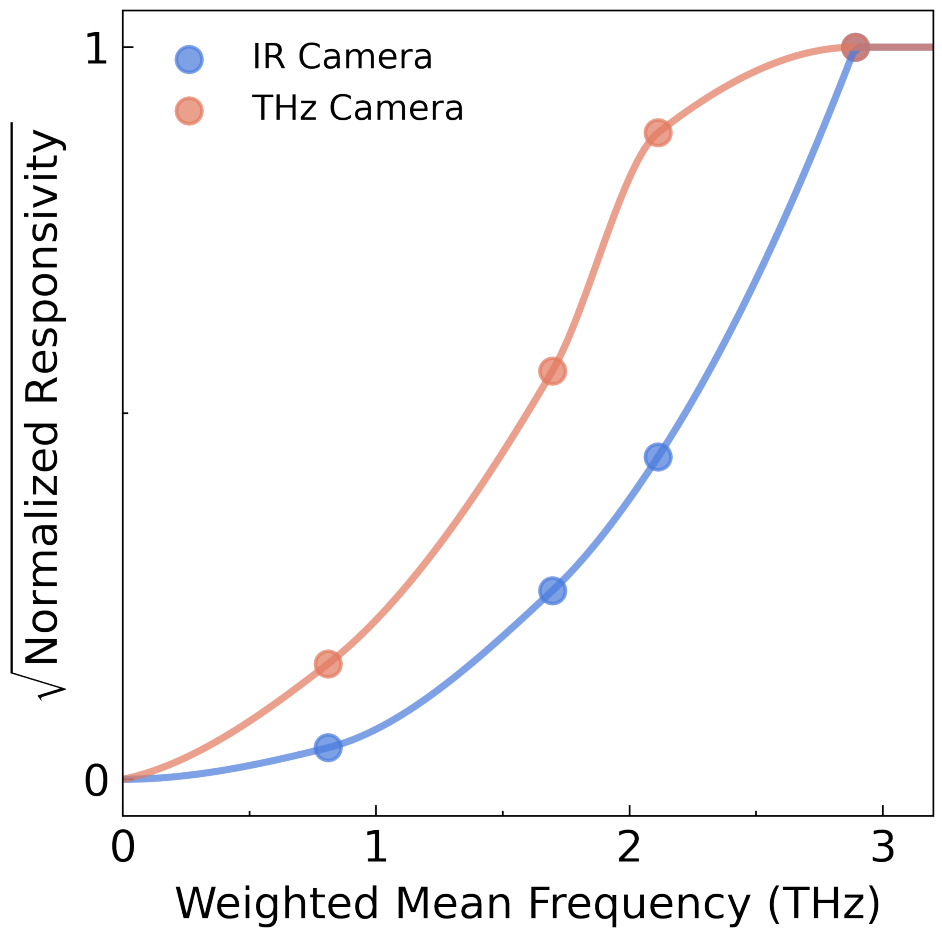

***Figure S3. Square root of the relative responsivity and interpolated curves.*** *The data were extracted for the IR (blue) and THz (red) cameras using the procedure described in section S1.*

The responsivity-weighted electric-field spectrum was then used to simulate the spatial distribution of the field intensity at the focus of the parabolic mirror. The simulation assumes that each spectral component propagates as a Gaussian beam, and the final beam profile will be the sum of the profiles from each spectral component. The Gaussian beam waist at the focus of the parabola is given by:

$$w(\nu) = \frac{cf}{\nu\pi w_{\mathrm{in}}}, \qquad \text{S2}$$

where $\nu$, $c$, $f$, $w_{in}$, are the THz frequency, speed of light, focal distance, and beam waist of the beam before the parabolic mirror. The corresponding intensity profile of each spectral component is given by:

$$I(\nu, r) = \frac{2|E(\nu)|^2}{\pi w(\nu)^2} exp\left(-\frac{2r^2}{w(\nu)^2}\right) \qquad \text{S3}$$

Where $E(\nu)$ is the responsivity weighted electric field, and $r = \sqrt{x^2 + y^2}$. The full detected profile is then the integral over all spectral components:

$$I_{\mathrm{total}}(x, y) = \int I\left(\nu, \sqrt{x^2 + y^2}\right) d\nu. \qquad \text{S4}$$

This procedure was implemented numerically using a 2D spatial grid and trapezoidal integration over the frequency axis.

## S4. Validation of the THz Origin of the Detected Signal

Microbolometric THz and IR cameras are sensitive to infrared radiation. Therefore, when it is used for THz beam profiling, particular care must be taken to ensure that the THz radiation is isolated from laser-induced thermal emission or residual pump leakage. Optical components exposed to the high-power femtosecond pump beam may heat up and emit infrared radiation, which can be detected by microbolometric cameras. Effective isolation from such thermal infrared backgrounds, through adequate THz filters, is therefore essential for reliable THz beam profiling. To verify that the recorded signal originates from THz radiation, we performed a series of control experiments, as described below.

**S4.1 Phase-Matching Dependence of the Detected Signal**

To confirm that the measured signal was indeed THz radiation generated by optical rectification, we placed a half-wave plate in front of the organic crystal and rotated the incoming (linear) polarization of the laser beam. When the polarization angle is not parallel with the organic crystal's optical axis, THz generation is suppressed, while other sources of radiation, such as laser-induced heating or pump leakage, are still present. Figure S4a shows horizontal cross-cuts of the intensity profile for different polarization angles θ (defined as the relative angle between the crystal axis and pump polarization), and Figures S4(b-e) show the respective beam profile. From these measurements, we can see that the signal vanishes for θ = 90°, consistent with THz generated by optical rectification.

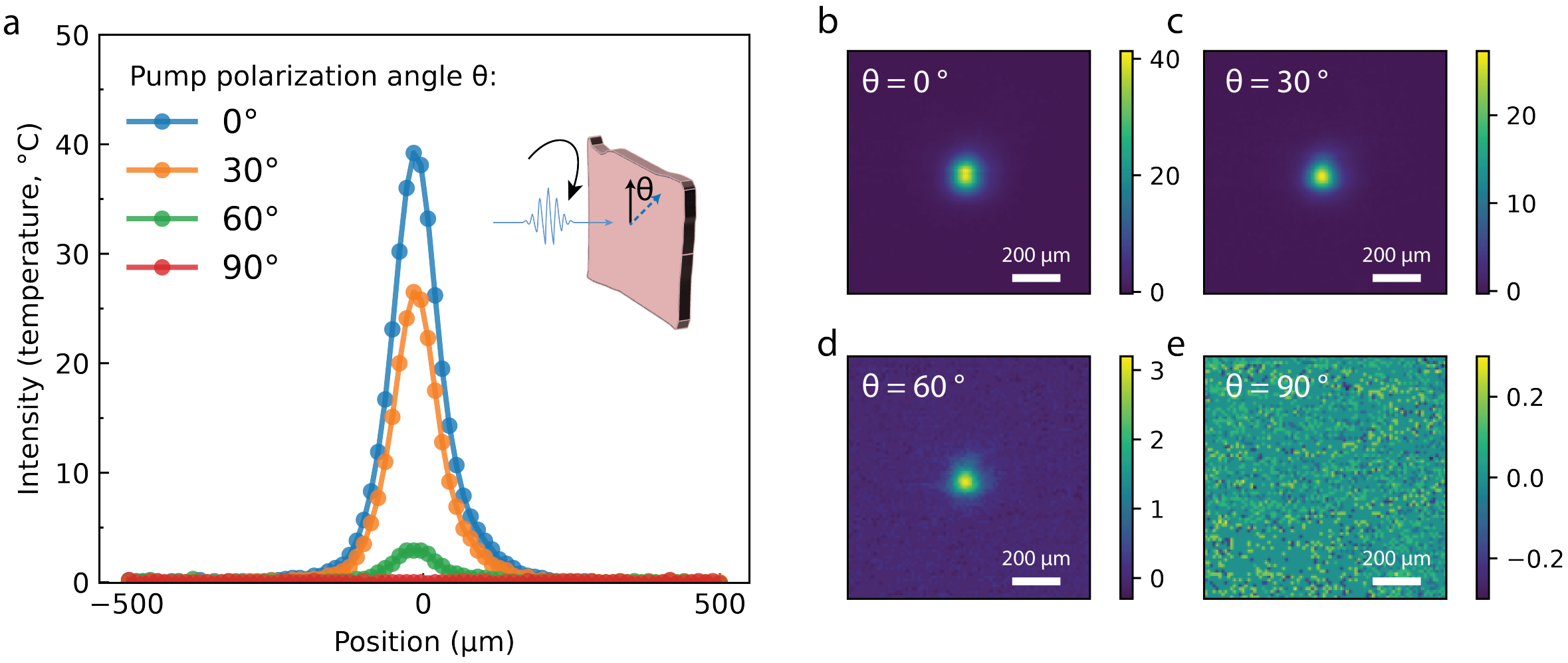

***Figure S4. Dependence of the detected signal on the pump polarization used for THz generation.*** *(a) Horizontal crosscuts of the intensity beam profiles recorded by the infrared microbolometric camera for different pump polarization angles (θ, defined as the relative angle between pump polarization and crystal axis). The inset schematically illustrates the rotation of the pump polarization relative to the organic crystal. The detected signal decreases strongly with increasing θ, reflecting the polarization dependence of the optical rectification process responsible for THz generation. (b–e) Corresponding beam images were recorded for pump polarization angles of 0°, 30°, 60°, and 90°, respectively. The color scale bars indicate the apparent temperature reported by the infrared camera. The beam intensity decreases as the pump polarization is rotated away from the optimal crystal orientation and disappears at 90°, where THz generation is suppressed.*

### S4.2 Suppression of the Signal by Removing the THz Source

Figures S5a,b show the optical setups used to verify the amount of signal measured from sources other than THz. Figure S5c shows the total acquired signal after pumping the organic crystal with OPA. Figure S5d shows the acquired images with the removal of the organic crystal. We observe that the signal drops below the noise floor, confirming that signals originating from thermal emission from the filters and downstream elements are completely suppressed. Figure S5e shows the horizontal crosscuts of the intensity profiles for the two configurations. These measurements confirm that laser-induced heating and pump leakage do not contribute to the acquired signals.

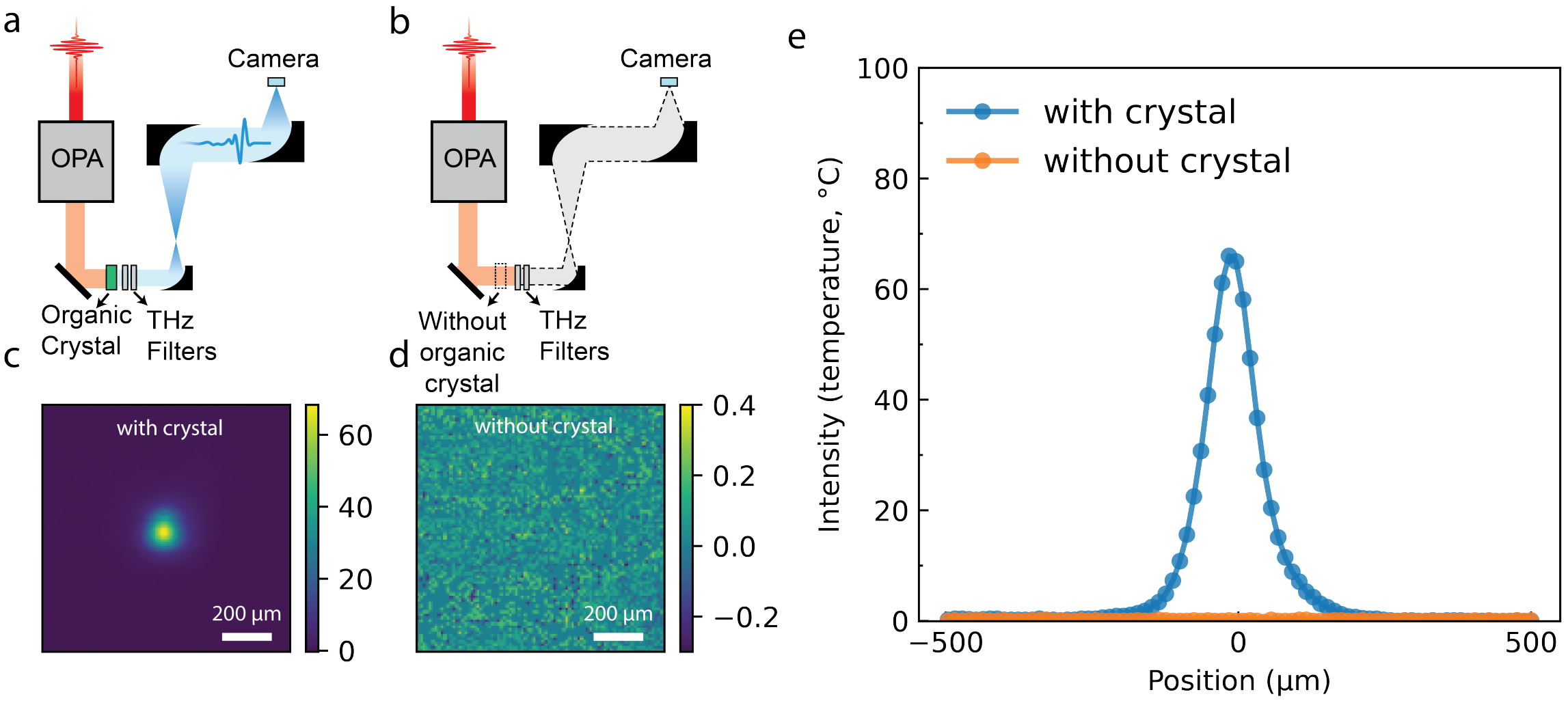

***Figure S5. Control experiments excluding infrared thermal radiation or pump leakage as the origin of the detected signal.*** *(a,b) Schematics of the experimental configurations used for the control measurements: (a) standard THz generation and detection configuration, and (b) configuration with the organic crystal removed, preventing THz generation. (c, d) Corresponding images recorded by the infrared microbolometric camera. A well-defined beam profile is observed only in the standard THz generation configuration (c), whereas no measurable signal is detected when the organic crystal is removed (d). The color scale bars indicate the apparent temperature reported by the infrared camera. (e) Horizontal cross-sections of the measured beam profiles for the three configurations. The disappearance of the signal when either THz generation is suppressed confirms that the detected response originates from THz radiation rather than from residual thermal emission from the filters.*

### S4.3 Characterizing the performance of the set of multi-mesh THz low-pass filters against a Teflon slab

All the experiments in this work were performed by isolating the optically rectified beam after the organic crystal with a set composed of two 20 THz and one 9 THz low-pass multi-mesh filters. The transmission spectra of these filters, measured by FTIR, are shown in Fig. S6. This combination of filters produces attenuation by more than 6 orders of magnitude across most of the mid- and near-IR spectral regions (calculated by multiplying each individual filter's transmission). For 1350 nm (pump wavelength), the attenuation ratio is on the order of $10^{-8}$. Under the conservative assumption that the full pump beam radiates on the filter stack, the transmitted power would be reduced to the nW–pW range. Even if this residual power were focused onto an area of approximately $\sim 100 \times 100\ \mu m$, corresponding to roughly 100 image pixels in our measurement, the power per pixel would remain in the pW–fW range, below the level required to produce a detectable signal.

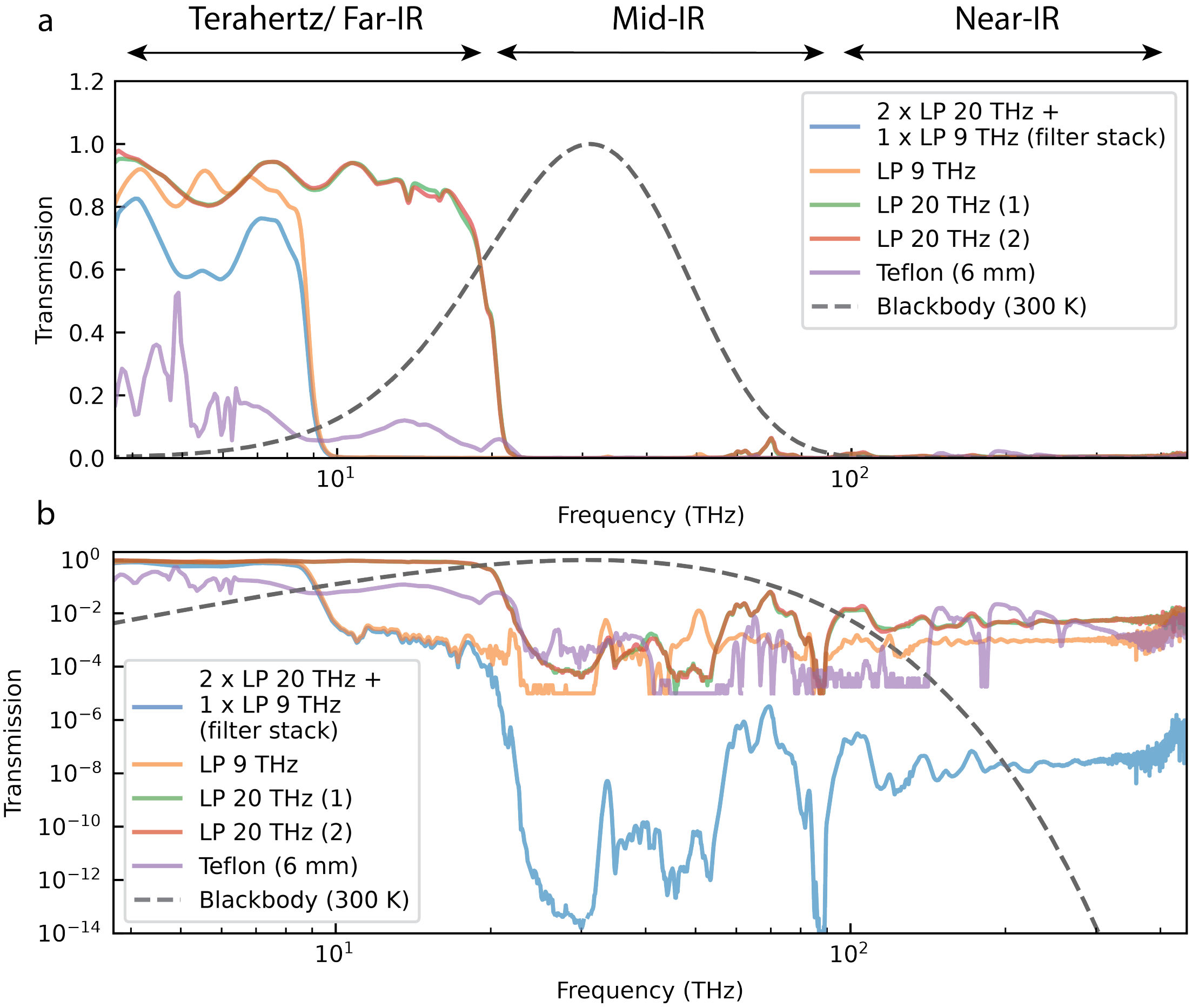

***Figure S6. Transmission characteristics of the THz filter stack and a PTFE slab.*** *(a) Linear-scale and (b) logarithmic-scale transmission spectra measured by FTIR for the individual QMC low-pass filters (LP 20 THz and LP 9 THz), the complete filter stack used in this work (two LP 20 THz filters and one LP 9 THz filter), and a 6-mm-thick PTFE (Teflon) slab. The complete filter stack was calculated by multiplying the transmission of individual filters. The dashed gray curve shows the normalized spectral radiance of a 300 K blackbody calculated using Planck's law.*

Fig. S6 also shows the blackbody radiation spectrum, estimated using Planck's law at 300 K. The spectral region where the blackbody radiation is most intense overlaps strongly with the spectral range rejected by the filter stack. Consequently, thermal infrared radiation generated by laser-induced heating of the organic crystal or surrounding optical components is expected to be strongly suppressed before reaching the detector, whereas the generated THz radiation remains largely transmitted.

While the filter stack provides excellent rejection of infrared radiation, alternative filtering approaches are also commonly employed in THz experiments. In particular, single PTFE (Teflon) slabs are commonly used in THz laboratories as broadband THz attenuators and optical filters. To compare their performance with the filter stack employed in this work, Fig. S6 shows the FTIR transmission spectra of a 6-mm-thick PTFE slab and of the multi-mesh filter stack. It is found that the PTFE slab, even being significantly thicker, performs similarly to individual multi-mesh filters across the mid-IR range; however, it has much lower transmission in the THz range (around 30%).

It is important to note that THz beam profiling with the HIKMICRO camera requires suppression not only of infrared radiation originating from the THz source, but also of thermal infrared radiation emitted by the filtering optics themselves and other downstream components. Therefore, more than one filtering stage can be advantageous, as thermal radiation emitted by one component may be attenuated by subsequent filters before reaching the detector. To exemplify this effect, in Fig. S7 we compare a THz beam profiling experiment done with the set of multi-mesh filters against a single 6 mm Teflon slab. While the multi-mesh filter stack yields a high-contrast beam profile with a low background level, the PTFE configuration exhibits a substantially elevated background signal surrounding the THz beam. This observation is consistent with residual thermal infrared radiation reaching the detector and illustrates the importance of cascading filtering stages when using infrared microbolometric cameras for THz beam profiling.

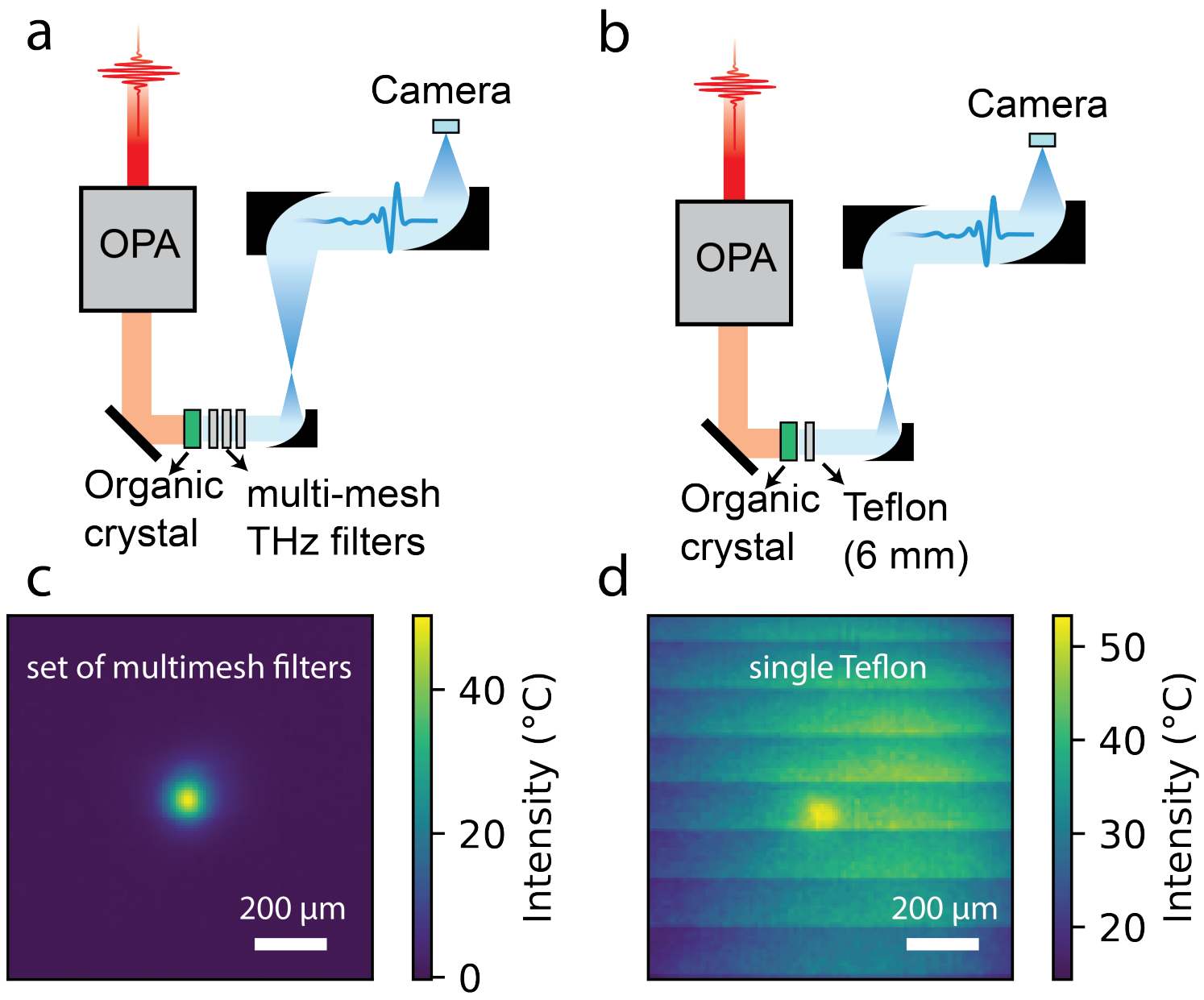

***Figure S7. Comparison between a multi-mesh filter stack and a single PTFE slab for THz beam profiling with an infrared microbolometric camera.*** *(a,b) Schematics of the experimental configurations employing (a) a stack of two 20 THz and one 9 THz low-pass multi-mesh filters and (b) a single 6-mm-thick PTFE (Teflon) slab. (c,d) Corresponding beam images recorded with the HIKMICRO camera. All experimental conditions were kept the same, except for the filtering components.*

### S4.4 Testing if the measured light is polarized

To further confirm the nature of the measured signal, in Figure S8a-e, we characterize the degree of polarization of the incoming beam. A THz polarizer was positioned in the beam path (Figure S8b), and the relative angle φ between the polarizer axis and the THz beam was varied. In Figure S8a, we show the horizontal crosscuts of the intensity profile, and in Figure S8c, we plot the integrated signal $I$ as a function of the polarizer angle φ. Fitting the integrated signal we obtained $I = (4490 \pm 80)\, cos^2[\varphi + (0.024 \pm 0.018)] + (33 \pm 65)$. Figures S8d and e show that the signal is completely suppressed when $\varphi$ = 90 degrees. These results confirm that the measured signal is linearly polarized, consistent with THz generated by optical rectification in the organic crystal. Thermal radiation, on the other hand, should be unpolarized. Figure S8f shows the transmission spectrum of the THz polarizer of unpolarized light across the entire THz-IR region.

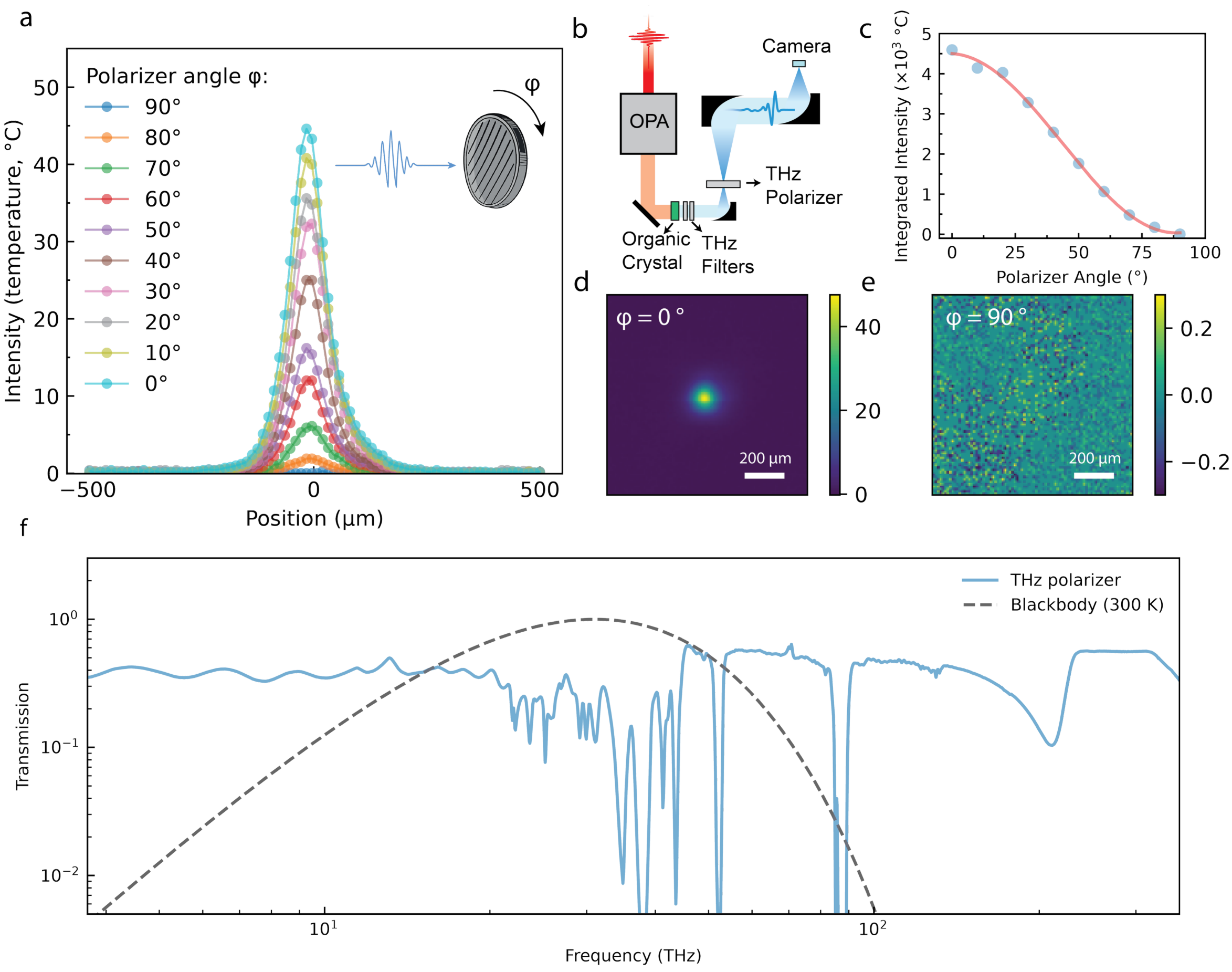

***Figure S8. Verification of the THz Origin of the Signal Through Polarization Measurements.*** *(a) Horizontal crosscuts of the intensity profiles recorded by the infrared microbolometric camera for different orientations of a THz polarizer placed in the beam path. The signal decreases continuously as the polarizer is rotated from 0° to 90°, at which point the transmitted THz radiation is strongly suppressed. Inset: Illustration of the THz polarizer used in the experiment, showing the rotation angle φ relative to the incident THz polarization. (b) Schematic of the experimental configuration used for the polarization-dependent measurements. A single THz polarizer is positioned in the beam path after the THz is generated. (c) Integrated beam intensity I as a function of polarizer angle φ, showing the expected polarization dependence of the transmitted THz field. The solid line is a fit to Malus's law. The fitting result is* $I = (4490 \pm 80)\,cos^2[\varphi + (0.024 \pm 0.018)] + (33 \pm 65)$ . *(d,e) Representative beam images recorded at polarizer angles of 0° and 90°, respectively. At 90°, the beam profile disappears, leaving only the detector noise floor. The color scale bars indicate the apparent temperature reported by the infrared camera. (f) Transmission of the THz polarizer (blue, solid line) and normalized black body radiation (Planck's law, grey, dashed line) spectra across the THz and infrared ranges, for unpolarized light.*

### S4.5 Effect of Ambient Humidity on the Detected Signal

To further confirm the origin of the detected signal, we performed measurements under different ambient humidity conditions. Water vapor exhibits strong absorption in the THz spectral range, whereas its effect on broadband thermal infrared radiation over the short propagation distances used in our experiment is comparatively small. As shown in Fig. S9, increasing the relative humidity to approximately 40% reduced the measured signal by about 32%. This behavior is consistent with attenuation of the THz beam due to water vapor absorption along the optical path. If the measured signal were primarily caused by thermal infrared emission from the source or surrounding components, a substantially weaker dependence on ambient humidity would be expected. The observed humidity dependence, therefore, provides further evidence that the signal detected by the IR microbolometric camera originates from incident THz radiation rather than from parasitic infrared thermal emission or pump leakage.

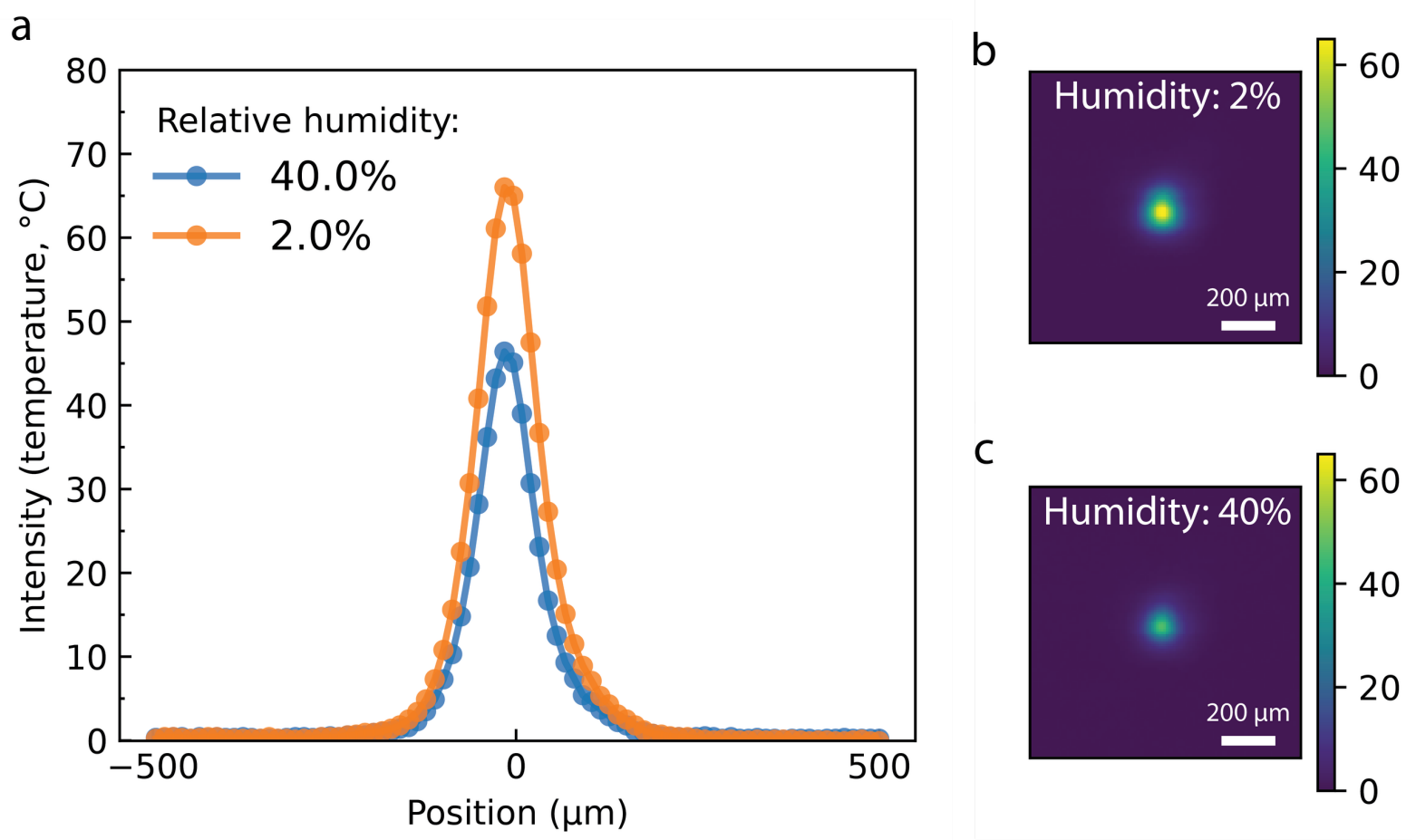

***Figure S9. Effect of ambient humidity on the THz signal detected by the infrared microbolometric camera.*** *(a) Horizontal cross-sections of the beam profile were measured at relative humidities of 2% and 40%. Increasing humidity reduces the peak signal by approximately 32%. (b,c) Corresponding THz beam images were recorded at relative humidities of 2% and 40%, respectively. The observed decrease in signal is consistent with attenuation of the THz radiation by atmospheric water vapor absorption. The color scale bars indicate the apparent temperature reported by the infrared camera.*

## Supplementary References